# Role of Two-Dimensional Ising Superconductivity in the Non-Equilibrium Quasiparticle Spin-to-Charge Conversion Efficiency

Kun-Rok Jeon,[*†] Kyungjune Cho,[†] Anirban Chakraborty, Jae-Chun Jeon, Jiho Yoon, Hyeon Han, Jae-Keun Kim and Stuart S. P. Parkin[*]

*Max Planck Institute of Microstructure Physics, Weinberg 2, 06120 Halle (Saale), Germany*

[†]These authors contributed equally to this work.

*To whom correspondence should be addressed: jeonkunrok@gmail.com,

stuart.parkin@halle-mpi.mpg.de

**ABSTRACT**

**Non-equilibrium studies of two-dimensional (2D) superconductors (SCs) with Ising spin-orbit coupling are prerequisite for their successful application to equilibrium spin-triplet Cooper pairs and, potentially, Majorana fermions. By taking advantage of the recent discoveries of 2D SCs and their compatibility with any other materials, we fabricate here non-local magnon devices to examine how such 2D Ising superconductivity affects the conversion efficiency of magnon spin to quasiparticle charge in superconducting flakes of 2H-NbSe$_2$ transferred onto ferrimagnetic insulating Y$_3$Fe$_5$O$_{12}$. Comparison with a reference device based on a conventionally paired superconductor shows that the Y$_3$Fe$_5$O$_{12}$-induced in-plane (IP) exchange spin-splitting in the NbSe$_2$ flake is hindered by its inherent out-of-plane (OOP) spin-orbit-field, which, in turn, limits the transition-state enhancement of the spin-to-charge conversion efficiency. Our out-of-equilibrium study highlights the significance of symmetry matching between underlying Cooper pairs and exchange-induced spin-splitting for the giant transition-state spin-to-charge conversion and may have implications towards proximity-engineered spin-polarized triplet pairing *via* tuning the relative strength of IP exchange and OOP spin-orbit fields in ferromagnetic**

insulator/2D Ising SC bilayers.

Injection and excitation of electrons, typically called Bogoliubov quasiparticles (QPs), in a superconductor (SC) with either external (Zeeman) or internal (exchange) spin-splitting field[1-3] under *non-equilibrium* conditions (*i.e.* voltage bias or temperature gradient) have been one of the central research topics in superconducting spintronics.[1-7] This is because their exotic transport properties, derived from the superconductivity-facilitated coupling between different non-equilibrium imbalances (*e.g.* spin, charge, heat and spin-heat), can considerably improve the functionality and performance of spintronic devices. Various non-equilibrium phenomena mediated by QPs have been observed in SC-based devices with either Zeeman or exchange spin-splitting: long-range spin signals,[8-10] pure thermal spin currents,[11] large (spin-dependent) thermoelectric currents[12] and spectroscopic evidence of spin-heat transport.[13]

Recently, a magnon spin-transport experiment[14] has reported that the conversion efficiency of thermal-magnon spin to QP charge *via* an inverse spin-Hall effect (iSHE)[15] in an exchange-spin-split Nb layer can be significantly enhanced by up to three orders of magnitude in the normal-to-superconducting transition regime. This giant transition-state QP iSHE has been semi-quantitatively explained in terms of two competing mechanisms of the superconducting coherence *versus* the exchange-field-frozen QP relaxation. A very recent theory[16] has pointed out that the electron-hole symmetry breaking present in SC/FMI (FMI = ferromagnetic insulator) bilayers mixes the spin and heat imbalances and can cause the enhancement of QP spin accumulation by several orders of magnitude relative to the normal state. Both these studies[14,15] emphasize the crucial role of the spin-splitting of QP density-of-states (DOS) and the resulting electron-hole asymmetry in enhancing the spin sensitivity of the SC detector.[5,15]

The advent of two-dimensional (2D) SCs[17-21] and their compatibility with any other materials *via* circumventing the need for lattice matching between adjacent material systems provide platforms to explore intriguing physical phenomena in various geometries,[22] including van der Waals (vdW) heterostructures with a twist, and in proximity combination with magnetic vdW flakes and/or thin films.[23,24] Because excited QPs and Cooper pairs in the superconducting condensate state are intimately correlated,[1-6] studies of non-equilibrium QP spin properties in such 2D SCs are of fundamental importance for understanding *equilibrium* spin-polarized triplet Cooper pairing[1-6] and the possible stabilization of Majorana fermions.[25-27]

2D superconductivity has been recently discovered in monolayer transition metal dicalcogenides (TMDs)[17] such as gated 2H-MoS$_2$[18,19] and 2H-NbSe$_2$.[20] Interestingly, the in-plane (IP) upper critical field $\mu_0 H_{c2}^{\parallel}$ is found to far exceed the Pauli paramagnetic limit of isotropic Bardeen-Cooper-Schrieffer (BCS) SCs $\mu_0 H_P^{BCS} \approx 1.84 T_c$,[28] where Zeeman spin-splitting fields are the predominant mechanism for Cooper pair breaking in the 2D limit and $T_c$ is the superconducting transition temperature. Such an enhancement of $\mu_0 H_{c2}^{\parallel}$ is explained by Ising spin-orbit coupling (SOC),[17-21] rooted in the broken IP crystal inversion symmetry plus the large SOC due to heavy transition metal atoms in TMDs. The Ising SO field $\mu_0 H_{SO}$ (as large as several hundred Tesla in the monolayer limit)[17-21] strongly pins Cooper pair spins at *K* and *K'* points of the hexagonal Brillouin zone to opposite out-of-plane (OOP) directions over IP applied magnetic fields. This stabilizes OOP Cooper pairing and forms so-called Ising superconductivity.[17-21]

We here investigate how the 2D Ising superconductivity influences the transition-state enhancement of magnon spin to QP charge conversion in a superconducting flake of 2H-NbSe$_2$[20,29-31] (Figure 1a) and compare its efficiency with a conventional superconducting thin film of Nb[14] (BCS SC). We firstly demonstrate that the normal-state spin-to-charge conversion functionality of the 2H-NbSe$_2$ flake can be *4 times more efficient* than that of the Nb film. We then find distinctively different transition-state conversion behaviours (*e.g.* modest transition-state enhancement, rather weak thickness dependence) in the 2H-NbSe$_2$ and attribute these to OOP Cooper pairing that hampers proximity penetration of IP exchange spin-splitting from the adjacent ferrimagnetic insulating Y$_3$Fe$_5$O$_{12}$. Notably, the maximum enhancement of spin-to-charge conversion appears at a critical thickness over which the IP crystal symmetry is recovered (equivalently, OOP Ising pairing is no longer protected), allowing the IP exchange field to penetrate. This provides a guideline as to how to tune the relative strength of these two phenomena for a desired proximity effect.[32,33] We believe that along with recent advances in 2D SCs of various intriguing properties (*e.g.* type-I/-II Ising, Rashba, topological SCs),[22,34] our approach helps find right material combinations for developing superconducting spintronic devices over conventional BCS SCs.

## RESULTS AND DISCUSSION

Our non-local magnon spin-transport devices (Figure 1a) are composed of two identical Pt electrodes and a central 2-H NbSe$_2$ flake transferred onto 200-nm-thick single-crystalline Y$_3$Fe$_5$O$_{12}$ (YIG) films (see Methods and Supplementary Section 1 for details), which are grown by liquid phase epitaxy on a (111)-oriented single-crystalline Gd$_3$Ga$_5$O$_{12}$, (GGG) wafer. Bulk 2H-NbSe$_2$ is a layered type-II SC, having anisotropy[29] in both the IP (OOP) coherence length $\xi_{SC}^{\parallel}$ ($\xi_{SC}^{\perp}$) ≈ 10 (3) nm and the IP (OOP) London penetration depth $\lambda_L^{\parallel}$ ($\lambda_L^{\perp}$) ≈ 70 (230) nm at zero temperature $T$ = 0. As shown in Figure 1b, it has a hexagonal crystal structure with lattice constants, $a = b$ ≈ 0.3 nm and $c$ ≈ 1.3 nm, and each unit cell consists of two AB stacked NbSe$_2$ layers.[30,31] On a single-piece YIG film, we prepare several independent devices with different 2H-NbSe$_2$ flake thicknesses $t_{NbSe_2}$ (Figure 1c-h), as well as reference devices in which Nb thin film is directly deposited[14] (Figure 1i,j). The Nb thickness $t_{Nb}$ is fixed at 15 nm, which is comparable to its dirty-limit coherence length $\xi_{Nb}$, so that the YIG-induced exchange spin-splitting-field can penetrate the Nb layer, while retaining the superconducting coherence, thereby maximizing the transition-state QP iSHE.[14]

In this device structure (Figure 1c,e,g,i), we pass a d.c. current $I_{dc}$ through one Pt electrode (using leads 1 and 2) while measuring the IP magnetic-field-angle $\alpha$ dependence of the non-local open-circuit voltages $[V_{nl}^{Pt}(\alpha), V_{nl}^{NbSe_2\ (or\ Nb)}(\alpha)]$ using the other Pt electrode (leads 7 and 8) and the central NbSe$_2$ (or Nb) (leads 3 and 4). Since we apply an external IP magnetic field $\mu_0 H_{ext}$ = 5 mT that is larger than the coercive field $\mu_0 H_c^{YIG}$ of YIG, $\alpha$ is simply defined as the relative angle of $\mu_0 H_{ext}$ (//$M_{YIG}$) to the long axis of the two Pt electrodes which are collinear.[14] The total voltage measured across the detector is then given by $V_{nl}^{tot} = \Delta V_{nl}^{el} + \Delta V_{nl}^{th} + V_0$. Here, $\Delta V_{nl}^{el}$ and $\Delta V_{nl}^{th}$ developed *via* iSHE (spin-to-charge conversion)[15] in the detector are proportional to the magnon spin current and accumulation created electrically [SHE (charge-to-spin conversion)[15] $\propto I_{dc}$], and thermally [spin-Seebeck effect (SSE, heat-to-spin conversion)[35] $\propto (I_{dc})^2$], respectively.[14,35] By inverting the polarity of $I_{dc}$, one can determine the magnitude of each component based on their characteristic angular dependences;[14,36] $\Delta V_{nl}^{el} = \frac{[V_{nl}^{tot}(+I_{dc}) - V_{nl}^{tot}(-I_{dc})]}{2} \propto \sin^2(\alpha)$ and $\Delta V_{nl}^{th} = \frac{[V_{nl}^{tot}(+I_{dc}) + V_{nl}^{tot}(-I_{dc})]}{2} - V_0 \propto \sin(\alpha)$. $V_0$ is a spin-independent offset voltage. Below, our discussion will focus on $\Delta V_{nl}^{th}$, since it remains detectably large at low $T$ for reasonable $|I_{dc}|$ such that Joule heating does not destroy the superconducting phase of the 2H-NbSe flake (or Nb thin film).

Let us first discuss the electrical transport properties of the transferred 2H-NbSe$_2$ flake. In the plot of its resistance $R^{NbSe_2}$ *versus* temperature $T$ (Figure 2a) for $t_{NbSe_2}$ = 9 nm, a resistance anomaly appears around 26 K, which is indicative of its phase transition from a normal metal to an incommensurate charge density wave (CDW) phase.[37] Note that the strongly suppressed CDW phase transition temperature, $T_{CDW}$ = 26 K for our $t_{NbSe_2}$ = 9 nm flake, is presumably due to the proximity coupling of the CDW with the magnetic order of YIG. In analogy with the Pauli effect[28] in conventional SCs, the Zeeman (or exchange) energy competes with the CDW condensation energy and hence $T_{CDW}$ is predicted to decrease in the presence of external (and/or internal) spin-splitting fields.[38] As $T$ is reduced further, 2H-NbSe₂ becomes superconducting below ~6.75 K. From the $T$-dependent upper critical field (Figure 2d), that is obtained by applying an external magnetic field either parallel $\mu_0 H^\parallel$ (Figure 2b) or perpendicular $\mu_0 H^\perp$ (Figure 2c) to the interface plane, we find $\xi^\parallel_{NbSe_2} \approx 8$ nm and $\xi^\perp_{NbSe_2} \approx 3$ nm using Ginzburg–Landau (GL) theory[39] (see Methods for a detailed discussion), so confirming the anisotropic superconducting state of 2H-NbSe₂.[29] The extrapolated value of $\mu_0 H^\parallel_{c2}$ at lower $T$ goes beyond $\mu_0 H^{BCS}_P$ = 12.4 T. Because the $t_{NbSe_2}$ = 9 nm flake corresponds to 7 × the unit cell and is much smaller than $\lambda^\perp_L \approx 230$ nm, neither IP crystal inversion symmetry or orbital effects (*i.e.* interlayer Meissner screening current) is fully recovered.[17] So Ising Cooper pairing[17-21] would account for the increase of $\mu_0 H^\parallel_{c2}$ over $\mu_0 H^{BCS}_P$. Note that a rather linear $\mu_0 H^\parallel_{c2}(T)$ behaviour for the intermediate $t_{NbSe_2}$ = 9 nm suggests that not only Ising SOC[20] but also Abrikosov vortex occupation[39] causes Cooper pair breaking (see Methods for details). These multiple characteristics are a measure of the high quality of our transferred 2H-NbSe₂ flake. In contrast, the deposited Nb thin film of $t_{Nb}$ = 15 nm has isotropic coherence lengths $\xi^\parallel_{Nb} \approx \xi^\perp_{Nb} \approx 12–13$ nm (Figure 2h) and its low-$T$ $\mu_0 H^\parallel_{c2}$ value is below $\mu_0 H^{BCS}_P$ = 8.3 T (Figure 2f-h), as would be expected from an isotropic BCS SC.

We now focus on how the conversion efficiency of magnon-carried spin to QP charge varies when the 2H-NbSe₂ becomes superconducting. Figure 3a,d,g shows the thermally driven non-local signal $[\Delta V^{th}_{nl}]^{NbSe_2}$ for the $t_{NbSe_2}$ = 4, 9 and 46 nm devices at various base temperatures $T_{base}$ around the superconducting transition $T_c$. In the normal state ($T_{base}/T_c > 1$), a negative $\left[\Delta V_{nl}^{th}\right]^{NbSe_2}$ (< 0) of a few tens of nanovolts is observed for $I_{dc}$ = |0.5| mA ($J_{dc}$ = |3.0| MA/cm$^2$). Given $\left[\Delta V_{nl}^{th}\right]^{Pt}$ > 0 (Supplementary Section 2) and $\left[\Delta V_{nl}^{th}\right]^{Nb}$ < 0 (Figure 3j), this indicates that the 4d heavy element Nb, having a negative spin-Hall angle $\theta_{SH}$ (< 0), governs spin-to-charge conversion characteristics in the normal-state 2H-NbSe$_2$. Upon entering the superconducting state ($T_{base}/T_c$ < 1), a clear enhancement of $\left[\Delta V_{nl}^{th}\right]^{NbSe_2}$ up to around a hundred nanovolts appears immediately below $T_c$ ($T_{base}/T_c \approx 0.99$) and then it decays towards zero, deep into the superconducting state. It is noteworthy that the for the normal state ($T_{base}$ > $T_c$), $\left[\Delta V_{nl}^{th}\right]^{NbSe_2}$ of the transferred 2H-NbSe$_2$ flakes go beyond $\left[\Delta V_{nl}^{th}\right]^{Nb}$ of the deposited Nb film, in particular, the $t_{NbSe_2}$ = 2.5 nm device reveals *4 times greater* signals (Supplementary Section 3), indicating high spin mixing conductance and spin transparency at the interface between our transferred 2H-NbSe$_2$ flakes and YIG film.

We systematically measure the $T_{base}$ dependence of the normalized $R^{NbSe_2}/R_{T=8\,K}^{NbSe_2}$ (Figure 3b,e,h) and $\left[\Delta V_{nl}^{th}\right]^{NbSe_2}$ (Figure 3c,f,i) with varying $I_{dc}$ in the Pt injector. The results are qualitatively similar to the magnon devices with Nb detectors[14] and also to the $t_{Nb}$ = 15 nm reference device studied here (Figure 3j-l). As $I_{dc}$ increases, $T_c$ of the 2H-NbSe$_2$ detector is progressively reduced (inset of Figure 3c,f,i) and the transition width broadens. As a result of this depressed superconductivity, caused by the combined effect of more populated spin-polarized QPs[5] and increased heat dissipation in the 2H-NbSe$_2$ at a high $I_{dc}$, a peak of the $\left[\Delta V_{nl}^{th}\right]^{NbSe_2}$ enhancement occurring in vicinity of $T_c$ (Figure 3c,f,i) shifts to a low $T_{base}$ and the enhancement regime widens. These demonstrate that the spin-to-charge conversion efficiency indeed rises when mediated by QPs in the transition state of 2H-NbSe$_2$/YIG bilayer, that is the enhanced spin-detection functionality of a 2D Ising SC in the normal-to-superconducting transition regime.

We next plot the normalized voltages $\left[\Delta V_{nl}^{th}\right]^{NbSe_2}/\left[\Delta V_{nl}^{th}\right]^{NbSe_2}_{T = 8\,K}$ (Figure 4a-c) and $\left[\Delta V_{nl}^{th}\right]^{Nb}/\left[\Delta V_{nl}^{th}\right]^{Nb}_{T = 8\,K}$ (Figure 4d) as a function of the normalized temperature $T_{\text{base}}/T_c$ for a quantitative analysis. With increasing $I_{dc}$, the peak amplitude strongly diminishes, the full-width-at-half-maximum (FWHM) broadens, and the peak position is away from $T_c$ (inset of Figure 4a-d). In addition to these generic features, one can find important quantitative differences between the 2H-NbSe$_2$ and Nb detectors[14] from the thickness dependence of the amplitude, FWHM and position (Figure 4f).

Firstly, the enhancement amplitude attained in the 2H-NbSe$_2$ detectors is relatively small $\left(\left[\Delta V_{nl}^{th}\right]^{NbSe_2}/\left[\Delta V_{nl}^{th}\right]^{NbSe_2}_{T = 8\,K} \leq 12\right)$ compared with the $t_{Nb}$ = 15 nm reference device with a similar lateral dimension, even though the 2H-NbSe$_2$ flakes (*e.g.* $t_{NbSe_2}$ = 4, 9 nm) possess a higher $T_c$ in thinner layers (Figure 4e). Secondly, the peak width and position abruptly change across 3 nm, coinciding with $\xi_{NbSe_2}^{\perp}$ (black vertical line in Figure 4e,f) below which thermal-fluctuation-enhanced $T_c$ suppression at the 2D limit is expected,[20,39] and they become almost $t_{NbSe_2}$-independent for thicker flakes. Note that the Nb dectectors[14] reveal a monotonic narrowing of FWHM and a peak shift closer to $T_c$ with increasing $t_{Nb}$. Thirdly, unlike the Nb detectors,[14] the maximum enhancement in the spin-to-charge conversion does not appear at $t_{NbSe_2} \approx \xi_{NbSe_2}^{\perp}$ and the $t_{NbSe_2}$-dependent enhancement is rather weak.

To account for these distinctively different conversion phenomena, we consider the layer thickness-dependent Ising superconductivity.[20,40] For a few monolayer 2H-NbSe$_2$, the IP crystal inversion symmetry is strongly broken by Se atoms (Figure 1b) and thus OOP Cooper pairing is protected and stabilized by the resulting Ising SO-field (76 meV in the monolayer limit).[20,41] In this regime, the YIG-induced IP exchange field (< 1 meV)[14,41] hardly spin-splits the QP DOS of the 2H-NbSe$_2$ and the transition-state enhancement of QP iSHE thus relies mostly on the superconducting-coherence-relevant resonant absorption,[14,16,42] leading to a modest enhancement. As the flake becomes thicker, the IP bulk crystal inversion symmetry is restored, which weakens the OOP Ising pairing and, in turn, enables the YIG-induced IP exchange field to propagate through. This explains why we obtain the maximum enhancement of the transition-state QP iSHE at $t_{NbSe_2} = 9$ nm ($> \xi^{\perp}_{NbSe_2}$). Note that, as a critical thickness value that is necessary to fully restore the IP bulk inversion symmetry (equivalently, to diminish Ising pairing) is larger than the coherence length, beyond this critical value, proximity extension of the YIG-induced IP exchange spin-splitting over the entire 2H-NbSe$_2$ layers is not very effective, limiting the enhancement amplitude. Furthermore, a Γ-centred Se-electron Fermi pocket, constituting a second band with a smaller superconducting gap, emerges in the 2H-NbSe$_2$ thicker than a few monolayers.[43] This second band whose gap energy seems weakly dependent of $t_{NbSe_2}$[43] can provide another path for spin-polarized QPs to enter the 2H-NbSe detector, effectively weakening the $t_{NbSe_2}$-dependent transition-state enhancement.

Our out-of-equilibrium study highlights the importance of symmetry matching between underlying Cooper pairs and exchange-induced spin-splitting for the giant transition-state enhancement of QP iSHE.[14,16] Based on this, we would predict a greater transition-state QP iSHE, for instance, in MnPS$_3$/NbSe$_2$ bilayers, where exchange spin-splitting[44] and SO fields are both OOP and thus match in the symmetry each other. Similarly, Rashba SC/YIG bilayers, where the Rashba SC has IP SO-fields,[34] would be another symmetry-matching combination. Our results may also provide a guideline for the proximity engineering of hybrid quantum materials that allow for exotic quantum phases (*e.g.* topological superconductivity with spin-polarized triplet pairs and/or Majorana zero modes)[25-27] at zero field in equilibrium.

## CONCLUSIONS

Our magnon spin-transport experiments with 2H-NbSe$_2$ detectors have shown that OOP Cooper pairing of Ising SC, derived by IP inversion symmetry breaking and strong SOC, hinders the proximity propagation of IP exchange spin-splitting, in turn limiting the transition-state enhancement of QP iSHE. Contrary to the magnon devices with Nb (BCS SC) detectors,[14] the maximum enhancement does not appear at $t_{NbSe_2} \approx \xi^{\perp}_{NbSe_2}$ but at a different critical thickness over which the IP crystal symmetry is recovered and so the OOP Ising pairing is no longer protected, allowing the IP exchange field to penetrate. This result should be taken into account for better proximity engineering of Ising SC triplet Josephson junctions with IP ferromagnets.[45] We believe that with the layer thickness-tunable OOP Cooper pairing[20,40] and IP exchange spin-splitting, 2D Ising SC/FMI bilayers have desirable material properties for the topological protection of spin-polarized triplet Cooper pairs[25] and Majorana fermions.[26,27] Our findings, together with recent progresses in 2D SCs and magnetic vdW crystals,[22,24] also raise the possibility of developing highly-efficient atomically-thin spin-to-charge converters *via* symmetry engineering.

## METHODS

**Device fabrication.** We fabricated the magnon spin-transport devices (Figure 1c,e,g,i) based on 200-nm-thick single-crystalline YIG films (from Matesy GmbH, https://www.matesy.de/en/products/materials/yig-single-crystal) as follows. We first defined a pair of Pt electrodes with an area of 1.5 × 50 μm$^2$, which were deposited by d.c. magnetron plasma sputtering at an Ar pressure of 4 × 10$^{-3}$ mbar These Pt electrodes are separated by a center-to-center distance $d^{Pt-Pt}$ of 15 μm, which is comparable to the magnon spin-diffusion length $l^m_{sd}$ estimated from our previous study.[14] For the reference device (Figure 1i), we defined the central 15-nm-thick Nb detector with a lateral dimension of 9 × 12 μm$^2$, which was grown by Ar-ion beam sputtering at a working pressure of 1.5 × 10$^{-4}$ mbar. Subsequently, we defined the outer Au(80 nm)/Ru(2 nm) leads and bonding pads, which were deposited by Ar-ion beam sputtering.

We next selected NbSe₂ flakes of suitable geometry and thickness, which were mechanically exfoliated from a high-quality single crystal (from HQ Graphene, http://www.hqgraphene.com/NbSe2.php) and first transferred onto SiO$_2$(300 nm)/Si substrates, *via* optical microscopy inspection. We then picked up the selected NbSe₂ flake and transferred it onto the central region of each magnon device (Figure 1c,e,g) using a polydimethylsiloxane-based dry transfer method (see Supplementary Section 1 for full details). All these processes have been conducted in an inert atmosphere glove box to prevent oxidation and degradation of the 2H-NbSe₂. Note that the 2H-NbSe₂ flakes and Nb thin film were prepared on the same-piece YIG film, confirming almost identical SHE/iSHE properties of the Pt injectors/detectors.

To prevent the unintentional contribution of iSHE from inner Au/Ru leads themselves to total voltage signals, we electrically isolate them from the active regime of magnon spin-transport by depositing a 10-nm-thick Al₂O₃ oxide layer in-between apart from the electrical contact parts on top of the central 2H-NbSe₂ (or Nb). Finally, we defined the inner Au(10 nm)/Ru(2 nm) leads, which were deposited by Ar-ion beam sputtering. Before depositing the inner Au/Ru leads, the NbSe₂ (or Nb) and Pt surface were gently Ar-ion beam etched for transparent electrical contacts between them.

**Superconducting transition measurement.** To characterize superconducting properties, d.c. electrical transport measurements were conducted on either transferred NbSe₂ flakes or deposited Nb thin films of the fabricated magnon devices attached on either IP (Figure 2b) or OOP (Figure 2c) rotatable holder in a Quantum Design Physical Property Measurement System (PPMS). Using electrical leads 3–6 (Figure 1c,e,g,i) with a 4-probe configuration, we measured the resistance $R$ *versus* temperature $T$ curves at the applied current $I \leq 10$ µA while decreasing $T$. The $T$-dependent IP (OOP) upper critical field $\mu_0 H_{c2}^{\parallel}$ ($\mu_0 H_{c2}^{\perp}$) of Figure 2d (Figure 2g) was obtained by applying an external magnetic field $\mu_0 H^{\parallel}$ ($\mu_0 H^{\perp}$) parallel (perpendicular) to the interface plan. The $\mu_0 H_{c2}^{\parallel}(T)$ and $\mu_0 H_{c2}^{\perp}(T)$ values are determined from the point where $R = 0.5 R_{T = 8\,K}$.

We estimated the $\xi_{NbSe_2}^{\parallel}$ and $\xi_{NbSe_2}^{\perp}$ values of the transferred 2H-NbSe$_2$ flake ($t_{NbSe_2} = $ 9 nm) from the $\mu_0 H_{c2}^{\parallel}(T)$ and $\mu_0 H_{c2}^{\perp}(T)$ data (Figure 2d), respectively, using an anisotropic GL theory[39] for $t_{NbSe_2} \gg \xi_{NbSe_2}^{\perp}$:

$$\mu_0 H_{c2}^{\parallel} = \frac{\Phi_0}{2\pi \xi_{NbSe_2}^{\parallel} \xi_{NbSe_2}^{\perp}} \left(1 - \frac{T}{T_c}\right), \quad (1a)$$

$$\mu_0 H_{c2}^{\perp} = \frac{\Phi_0}{2\pi \left(\xi_{NbSe_2}^{\parallel}\right)^2} \left(1 - \frac{T}{T_c}\right), \quad (1b)$$

where $\Phi_0 = \frac{h}{2e} = 2.07 \times 10^{-15}\,Tm^2$ is the magnetic flux quantum. It is noteworthy that as $t_{NbSe_2}$ is reduced and reaches the atomically-thin limit ($t_{NbSe_2} \ll \xi_{NbSe_2}^{\perp}$), the dominant Cooper-pair breaking mechanism under application of $\mu_0 H^{\parallel}$ changes from Abrikosov vortex occupation to Ising SOC as recently discussed.[17-21] For $t_{NbSe_2} \ll \xi_{NbSe_2}^{\perp}$, Eq. (1a) can thus be rewritten as

$$\mu_0 H_{c2}^{\parallel} = \sqrt{\mu_0 H_{SO}^{Ising} \mu_0 H_P^{BCS} \left(1 - \frac{T}{T_c}\right)}, \quad (1c)$$

where $\mu_0 H_{SO}^{Ising}$ is the strength of Ising SO field. For completeness, we also fitted the $\mu_0 H_{c2}^{\parallel}(T)$ data (violet solid line, Figure 2d) with this formula.

On the other hand, for the deposited Nb thin film of $t_{Nb} = 15$ nm $\leq \xi_{Nb}^{\perp} = \xi_{Nb}^{\parallel}$, the $T$-dependent upper critical fields (Figure 2h) were fitted with:[39]

$$\mu_0 H_{c2}^{\parallel} = \frac{\Phi_0 \sqrt{12}}{2\pi \xi_{Nb}^{\parallel} t_{Nb}} \sqrt{\left(1 - \frac{T}{T_c}\right)}, \quad (2a)$$

$$\mu_0 H_{c2}^{\perp} = \frac{\Phi_0}{2\pi \left(\xi_{Nb}^{\parallel}\right)^2} \left(1 - \frac{T}{T_c}\right). \quad (2b)$$

Note that unlike bulk Nb, the occupation energy of Abrikosov vortices in a superconducting Nb thin film ($t_{Nb} \leq \xi_{Nb}^{\perp} = \xi_{Nb}^{\parallel}$) under $\mu_0 H^{\parallel}$ is higher than that under $\mu_0 H^{\perp}$, differentiating formulas [Eqs. (2a) and (2b)] for the $T$-dependent IP/OOP upper critical fields.[39] This is because the density of Cooper pairs cannot change much on a length scale shorter than the coherence length and hence IP Abrikosov vortices cannot efficiently accommodate magnetic flux.[39] When the Nb (BCS SC) film becomes sufficiently thin ($t_{Nb} \ll \xi_{Nb}^{\perp} = \xi_{Nb}^{\parallel}$), Abrikosov vortex occupation under $\mu_0 H^{\parallel}$ is strongly suppressed and a $\mu_0 H^{\parallel}$-driven dominant Cooper-pair breaker is now the Pauli paramagnetic effect (*i.e.* Zeeman spin-splitting).[28] Accordingly, Eq. (2a) can be rewritten by

$$\mu_0 H_{c2}^{\parallel} = \mu_0 H_P^{BCS} \sqrt{\left(1 - \frac{T}{T_c}\right)}. \qquad (2c)$$

**Non-local measurements.** We measured the non-local magnon spin-transport (Figure 1a) on the magnon devices attached on an IP rotatable sample holder in the Quantum Design PPMS at various *T* between 2 and 300 K. A d.c. current $I_{dc}$ in the range of 0.1 to 1 mA was applied to the first Pt using a Keithley 6221 current source and the non-local voltages [$V_{nl}^{Pt}(\alpha)$, $V_{nl}^{NbSe_2 \ (or \ Nb)}(\alpha)$] across the second Pt and the central 2H-NbSe2 (or Nb) are simultaneously recorded as a function of IP magnetic-field-angle *α* with rotating the IP sample holder by a Keithley 2182A nanovoltmeter. Note that *α* is defined as the relative angle of $\mu_0 H_{ext}$ (//$M_{YIG}$) to the long axis of two Pt electrodes which are collinear.

The Oersted field $\mu_0 H_{Oe}$ induced from $I_{dc}$ applied to the Pt electrode is estimated using Ampere's law, $\mu_0 H_{Oe} = \frac{\mu_0 I_{dc}}{2\pi w_{Pt}} \ln\left[1 + \frac{w_{Pt}}{d}\right]$. Here $\mu_0 = 4\pi \times 10^{-7} \ Tm/A$ is the permeability of free space, $w_{Pt}$ is the width (1.5 μm) of the Pt electrode and *d* is the distance from the Pt/YIG interface. For the maximum $I_{dc}$ = 1.0 mA used, we get $\mu_0 H_{Oe}$ = 0.3−0.4 mT at *d* = 100 nm and it decreases to 0.02−0.03 mT at *d* = 7.5 μm. These estimated values are *too weak* to perturb the magnetization direction of ferrimagnetic insulating YIG[14] under application of $\mu_0 H^{\parallel}$ = 5 mT (Figure 1c,e,g,i) and to suppress the superconducting properties of 2H-NbSe2 flakes and a Nb thin film whose upper critical fields in the transition state are larger than 0.5 T (Figure 2b,c,f,g).

## ASSOCIATED CONTENT

**Supporting Information**

Supporting Information for this article is available at [TBD].

Dry transfer of 2H-NbSe$_2$ flakes onto magnon spin-transport devices, Non-local spin signals detected by the Pt detector across $T_c$ of the 2H-NbSe$_2$ flake, Transition-state enhancement of QP iSHE for the $t_{NbSe_2}$ = 2.5 nm device.

## AUTHOR INFORMATION

**Author Contributions**

K.-R.J. and S.P.P.P. conceived and designed the experiments. The magnon spin-transport devices were fabricated by K.-R.J. with help from J.Y., J.-C.J., A.C., H.H., J.-K.K. and K.C. K.-R.J. performed exfoliation/pick-up/transfer of 2H-NbSe$_2$ under the guidance of K.C. The non-local transport measurements were carried out by K.-R.J. with help of J.Y. and J.-C.J. K.-R.J. performed the data analysis. S.P.P.P. supervised the project. All authors discussed the results and commented on the manuscript, which was written by K.-R.J., K.C. and S.S.P.P.

**Competing Interests**

The authors declare no competing financial interests.

## ACKNOWLEDGMENTS

This work was supported by the Alexander von Humboldt Foundation.

**FIGURE LEGENDS**

**Figure 1. Nonlocal magnon spin-transport device with Ising superconductor. a**) Device layout and measurement scheme. When a d.c. charge current $I_{dc}$ is applied to the right Pt injector, either electrically or thermally driven magnons accumulate in the ferrimagnetic insulator $Y_3Fe_5O_{12}$ (YIG) underneath and diffuse toward the left Pt detector. These magnon ($s = +1$) currents are then absorbed by the left Pt detector, resulting in the electron spin accumulation that is, in turn, converted to a nonlocal charge voltage $V_{nl}^{Pt}$ via the inverse spin-Hall effect (iSHE). Such a conversion process also occurs for the central 2H-NbSe$_2$ flake and thereby $V_{nl}^{NbSe2}$. Note that unlike spin-singlet ($S = 0$) Cooper pairs in a coherent ground state, the excited quasiparticles (QPs) can carry spin angular momentum in the superconducting state. How out-of-plane (OOP) Cooper pairing of the 2H-NbSe$_2$ affects the transition-state enhancement of QP iSHE will be discussed in this study. **b**) Crystal structure of the 2H-NbSe$_2$, where in-plane inversion symmetry breaking by Se plus spin-orbit coupling of Nb lead to OOP spin-singlet ($S = 0$) Cooper pairs, constituting Ising superconductivity. **c,e,g,i**) Optical micrographs of the fabricated devices. Atomic force microscopy (AFM) scans of the transferred 2H-NbSe$_2$ flakes (**d,f,h**) and the deposited Nb thin film (**j**).

**Figure 2. Electrical characterization of the transferred 2H-NbSe$_2$ flake. a**) 2H-NbSe$_2$ resistance $R^{NbSe2}$ as a function of temperature $T$ for the transferred 2H-NbSe$_2$ flake ($t_{NbSe_2}$ = 9 nm) measured using a 4-terminal current-voltage method (using leads 3, 4, 5, 6 in Figure 1e). Typical $R^{NbSe2}$-$T$ curves measured by applying an external magnetic field either parallel $\mu_0H^\parallel$

(**b**) or perpendicular $\mu_0 H^\perp$ (**c**) to the interface plane. The *T*-dependent IP (OOP) upper critical field $\mu_0 H_{c2}^\parallel$ ($\mu_0 H_{c2}^\perp$) is determined from the point where $R = 0.5 R_{T=8\,K}$. **d**) Summary of the $\mu_0 H_{c2}^\parallel(T)$ and $\mu_0 H_{c2}^\perp(T)$ data. The blue dashed line represents the Pauli paramagnetic limit $\mu_0 H_P^{BCS} \approx 1.84 T_c$.[28] The red and violet solid lines in **b** are theoretical fits using Ginzburg–Landau (GL)[39] and pair breaking (PB)[20] theories, respectively. **e,f,g,h**) Data equivalent to **a,b,c,d** but for the $t_{Nb} = 15$ nm reference device (Figure 1j).

**Figure 3. Enhancement of non-local signals in the transition state of the 2H-NbSe₂ detector. a,d,g**) Thermally driven non-local voltages $\left[\Delta V_{nl}^{th}(\alpha)\right]^{NbSe_2}$ as a function of IP field angle $\alpha$ for the $t_{NbSe_2} = 4$, 9 and 46 nm devices, taken at $I_{dc} = |0.5|$ mA around the superconducting transition $T_c$ of the 2H-NbSe₂. The black solid lines are $\sin(\alpha)$ fits. Note that dips in $\left[\Delta V_{nl}^{th}\right]^{NbSe_2}$ at $\alpha \approx 90°$ and 270° near $T_c$ which are pronounced for a thicker flake arise from Abrikosov-vortex-flow-driven *spin-independent* Hall effect[14] under a transverse magnetic field that is close to the upper critical field $\mu_0 H_{c2}$ of type-II SC (*i.e.* vortex melting field). **b,e,h**) Normalized 2H-NbSe₂ resistance $R^{NbSe_2}/R_{T=8\,K}^{NbSe_2}$ versus $T_{base}$ plots for the $t_{NbSe_2} = 4$, 9 and 46 nm devices, measured using a four-terminal current-voltage method (using leads 3, 4, 5, 6 in Figure 1c,e,g) with varying $I_{dc}$ in the Pt injector. The critical temperature $T_c$ is defined as the point where $R^{NbSe_2} = 0.5 R_{T=8\,K}^{NbSe_2}$. The inset summarizes the measured $T_c$ as a function of $I_{dc}$ (or $J_{dc}$). **c,f,i**) Estimated magnitude of $\left[\Delta V_{nl}^{th}\right]^{NbSe_2}$ as a function of $T_{base}$ for the $t_{NbSe_2} = 4$, 9 and 46 nm devices. **j-l**) Data equivalent to **a-c** but for the $t_{Nb} = 15$ nm reference device.

**Figure 4. 2H-NbSe₂ thickness dependence of the transition-state enhancement and comparison with the Nb detector. a-c**) $\left[\Delta V_{nl}^{th}\right]^{NbSe_2} / \left[\Delta V_{nl}^{th}\right]_{T=8\,K}^{NbSe_2}$ versus $T_{base}/T_c$ plot for the $t_{NbSe_2} = 4$, 9 and 46 nm devices. Each inset displays the $|I_{dc}|$ (or $|J_{dc}|$) dependence of the peak

amplitude, width, and position. **d**) Data equivalent to **a** but for the $t_{Nb} = 15$ nm device. Note that unlike the amplitude, the width and position can be approximately estimated based on data below $T_c$ (Figure 3c,f,i,l) where the transition-state enhancement of QP iSHE provides a detectable amplitude of $\left[\Delta V_{nl}^{th}\right]^{NbSe_2}$. **e**) $t_{NbSe_2}$-dependent $T_c$. **f**) $t_{NbSe_2}$-dependent peak amplitude, width and position. Abrupt changes of $T_c$, peak width and position below $t_{NbSe_2} = 3$ nm, coinciding with the OOP coherence length $\xi_{NbSe_2}^{\perp}$ (black vertical line in **e** and **f**), are likely due to thermal-fluctuation-enhanced $T_c$ suppression at the 2D limit.[20,39] Detailed results of the $t_{NbSe_2} = 2.5$ nm device can be found in Supplementary Section 3. In **e** and **f**, data from the $t_{Nb} = 15$ nm reference device are also included for quantitative comparison.

**a**

iSHE  
$V_{nl}^{Pt}$  
$V_{nl}^{NbSe_2}$  
Pt  
2H-NbSe$_2$  
SHE  
Pt  
$s = -1/2$  
$s = +1/2$  
$S = 0$  
$s = -1/2$  
$s = +1/2$  
Magnon spin-transport  
$s = +1$  
$\Delta T$  
YIG  
$M_{YIG}$  
GGG sub.  
$I_{dc}$

**b**

● Nb  
● Se  

1 H  
2 H  
vdW gap  
$a = b = 0.344$ nm, $c = 1.255$ nm  

$\alpha$  
$H_{ext} > H_c^{YIG}$

**c**  Pt, Al$_2$O$_3$, Transferred 2H-NbSe$_2$, YIG, $H_{ext}$, $\alpha$, $V_{nl}^{NbSe_2}$, $V_{nl}^{Pt}$, $I_{dc}$, 20 μm

**d** Height (nm) vs Distance (μm), 4 nm

**e**  Pt, Al$_2$O$_3$, Transferred 2H-NbSe$_2$, YIG, 20 μm

**f** Height (nm) vs Distance (μm), 9 nm

**g**  Pt, Al$_2$O$_3$, Transferred 2H-NbSe$_2$, YIG, 20 μm

**h** Height (nm) vs Distance (μm), 46 nm

**i**  Pt, Al$_2$O$_3$, Deposited Nb, YIG, $V_{nl}^{Nb}$, 20 μm

**j** Height (nm) vs Distance (μm), 15 nm

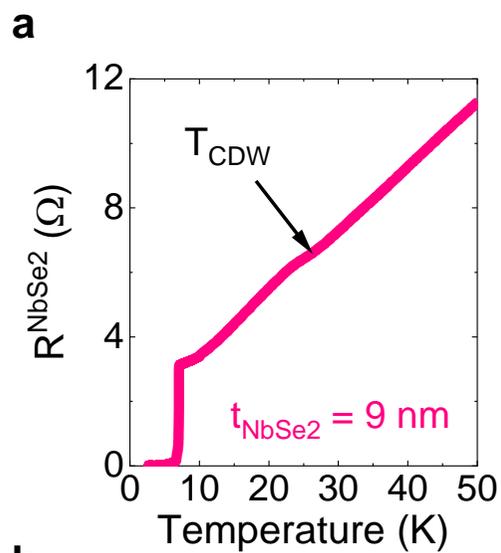
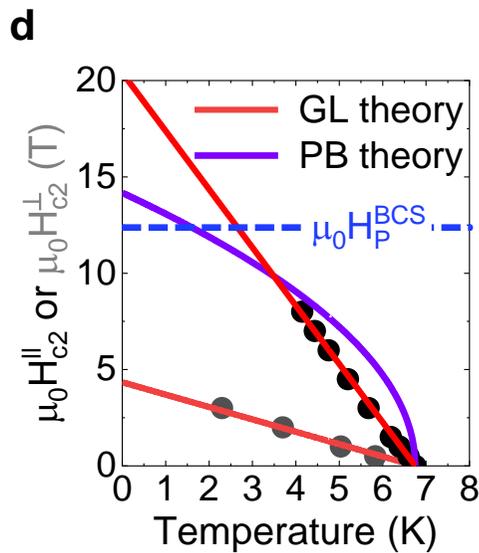
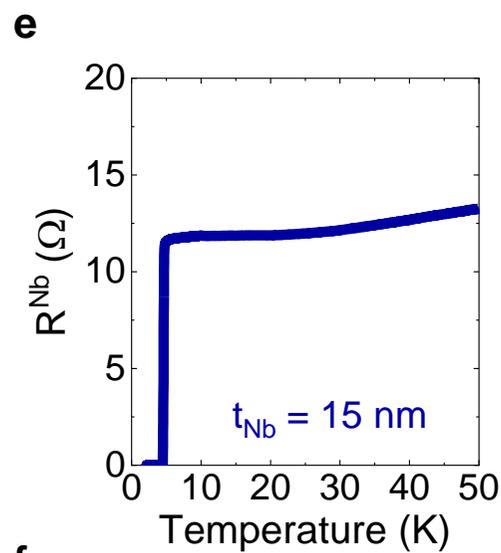
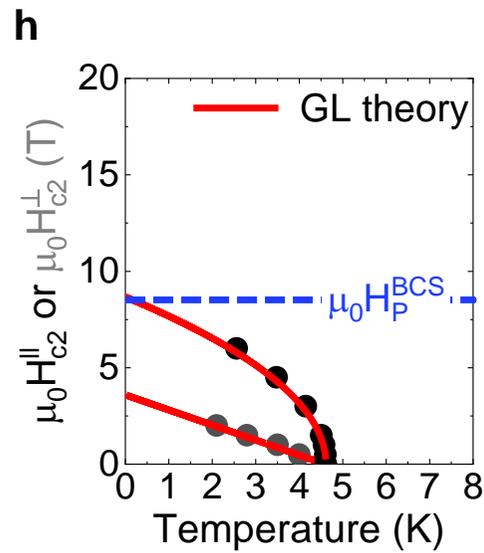
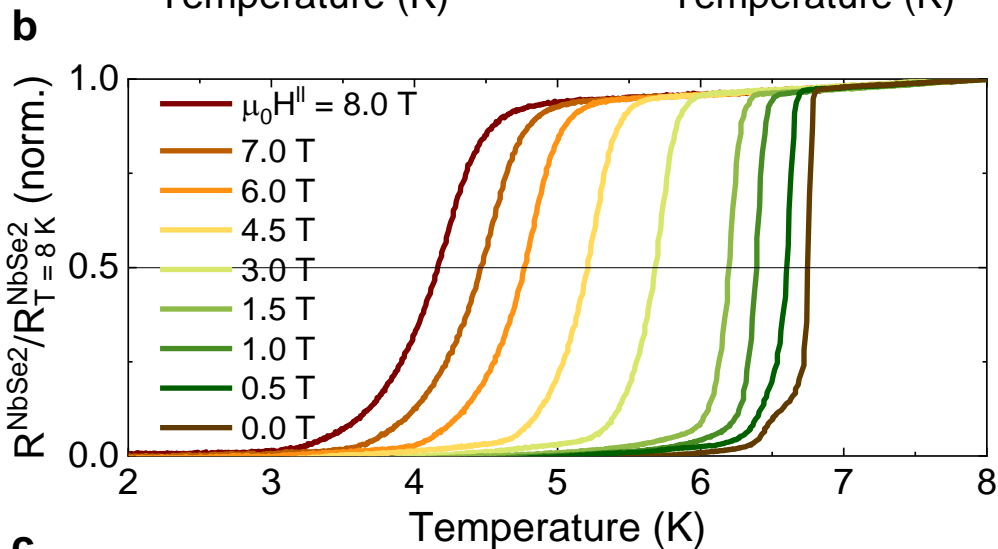
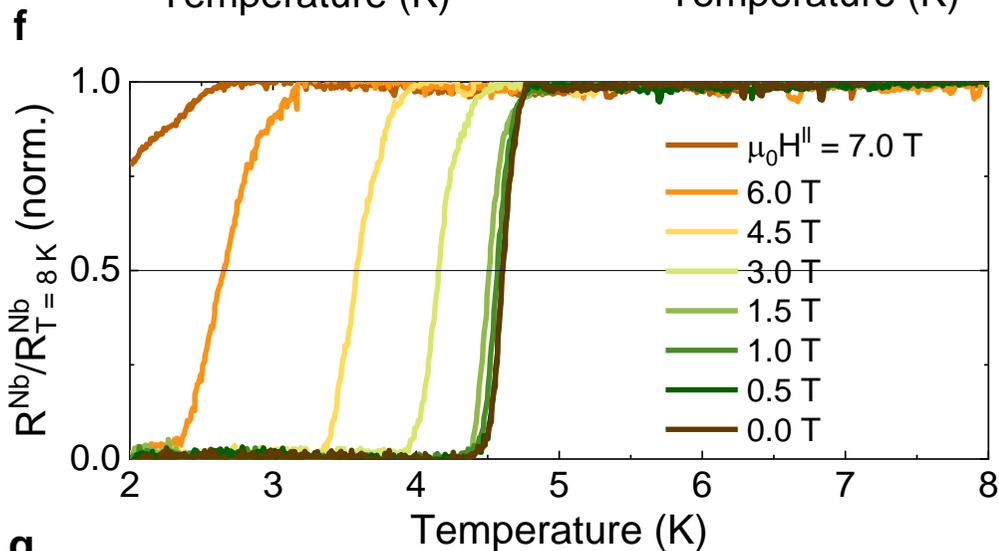
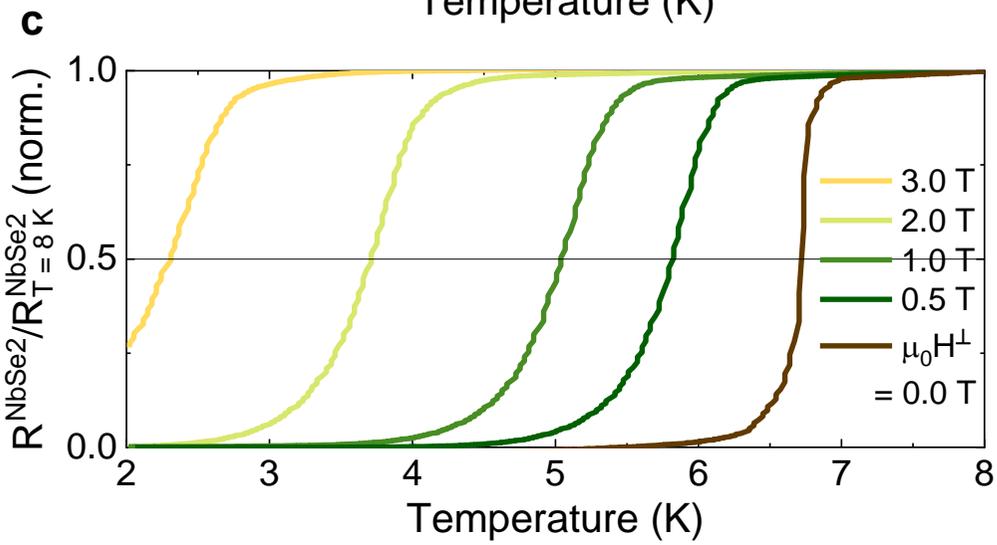
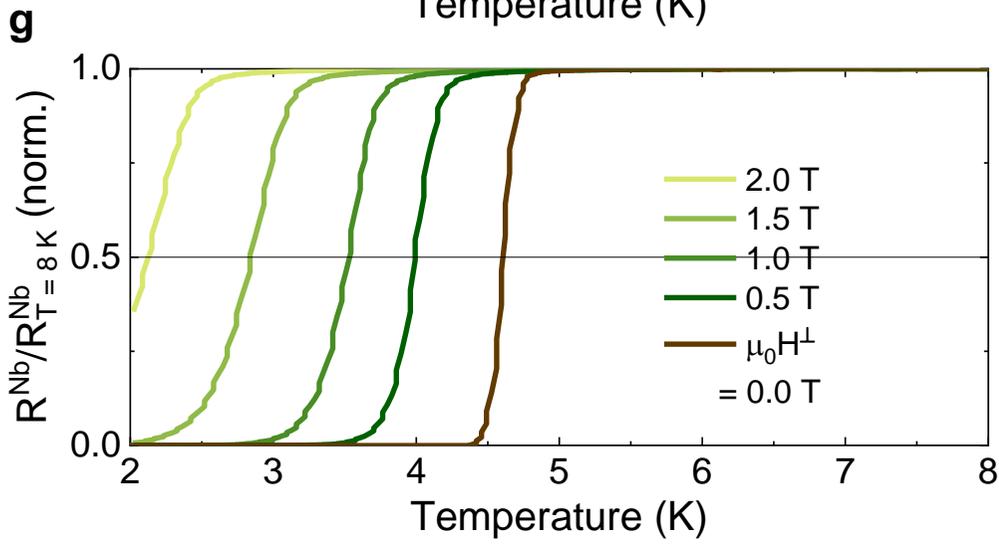

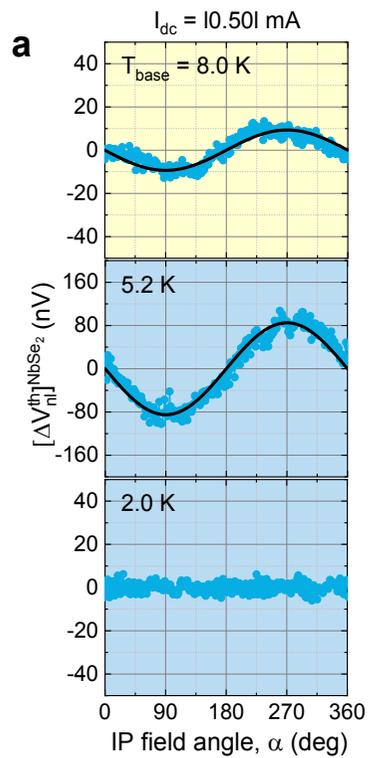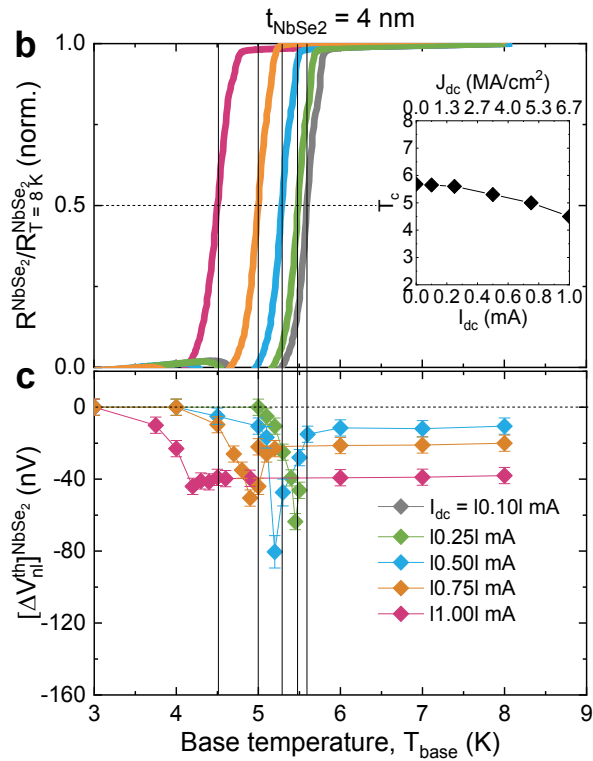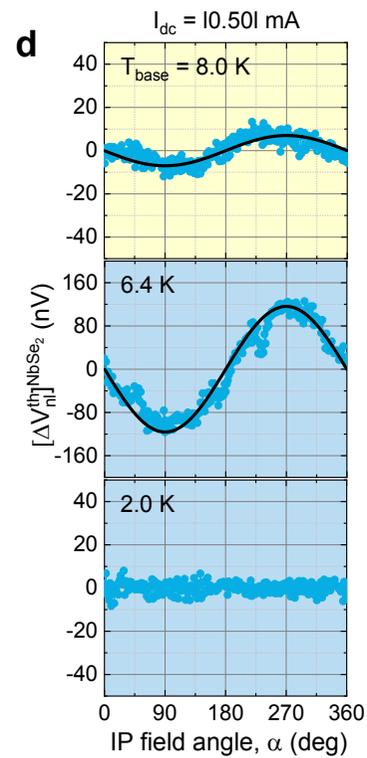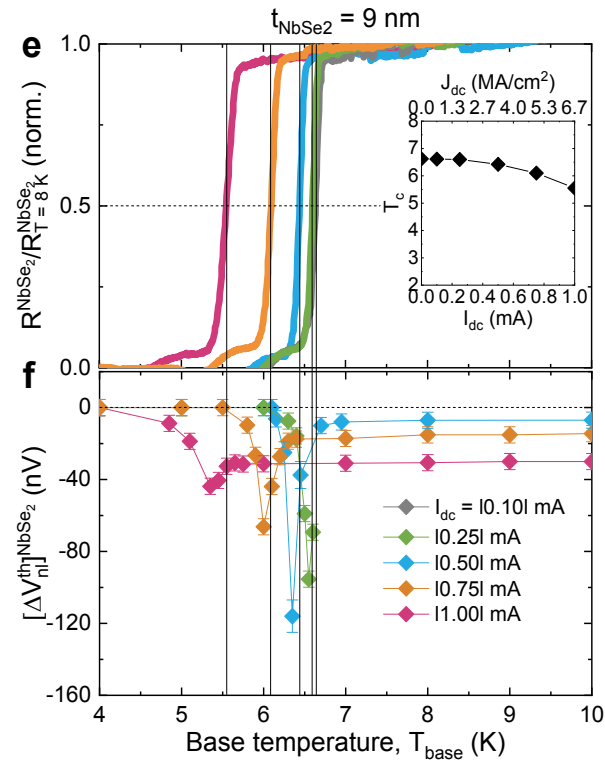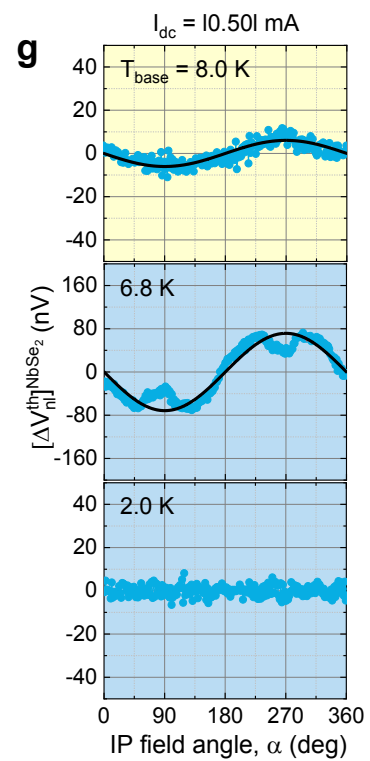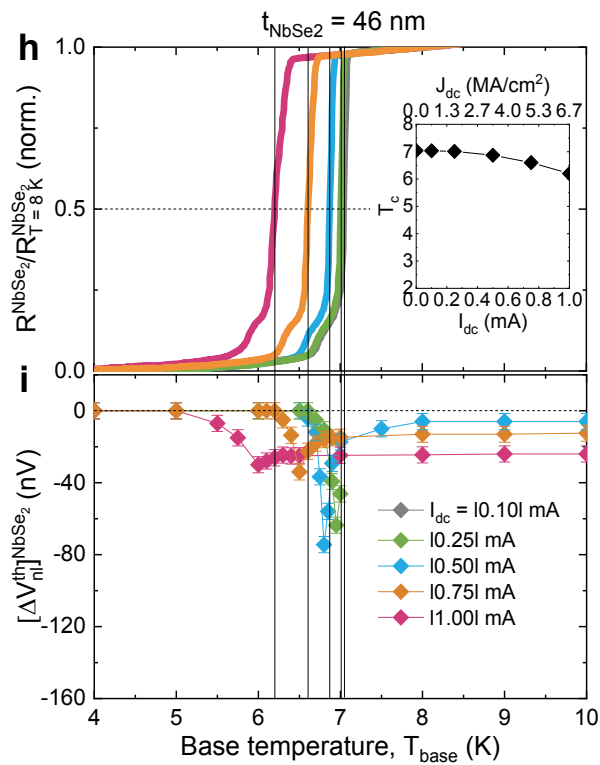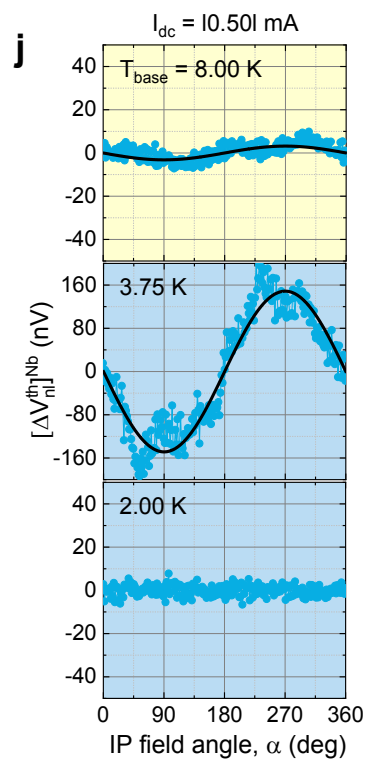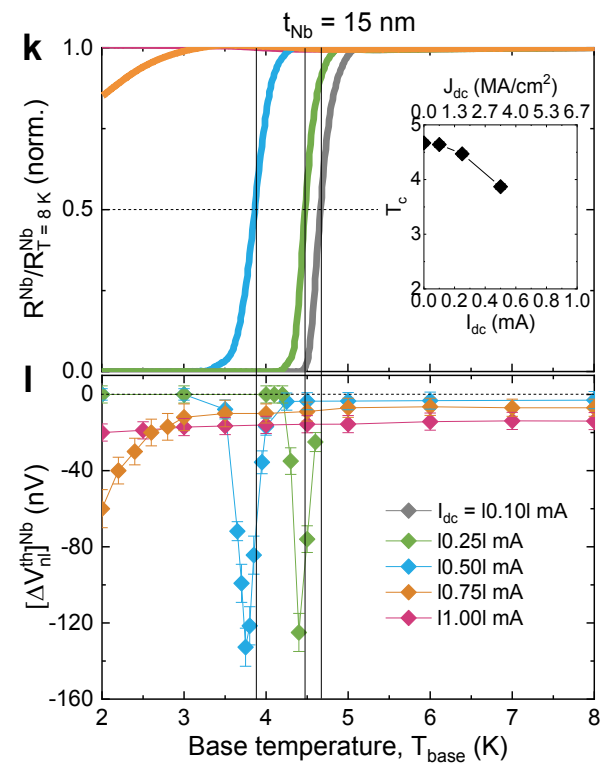

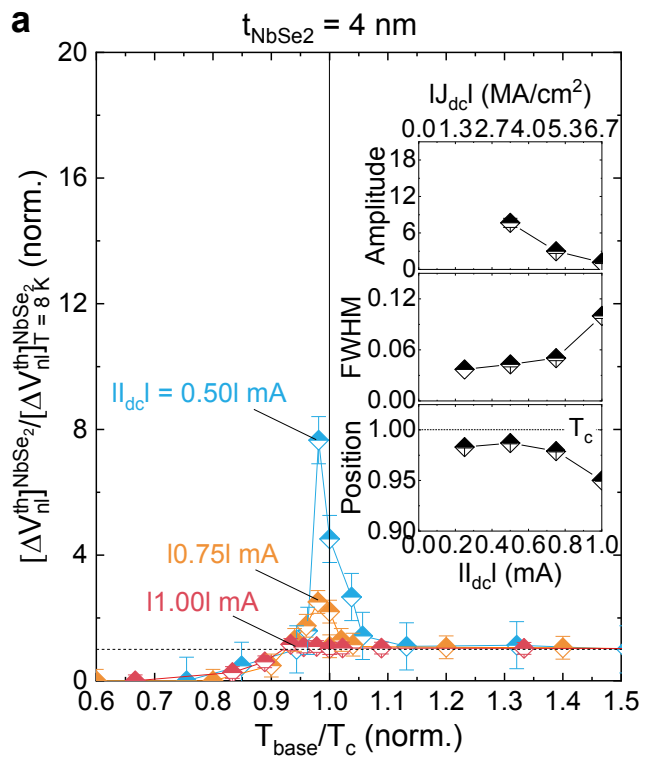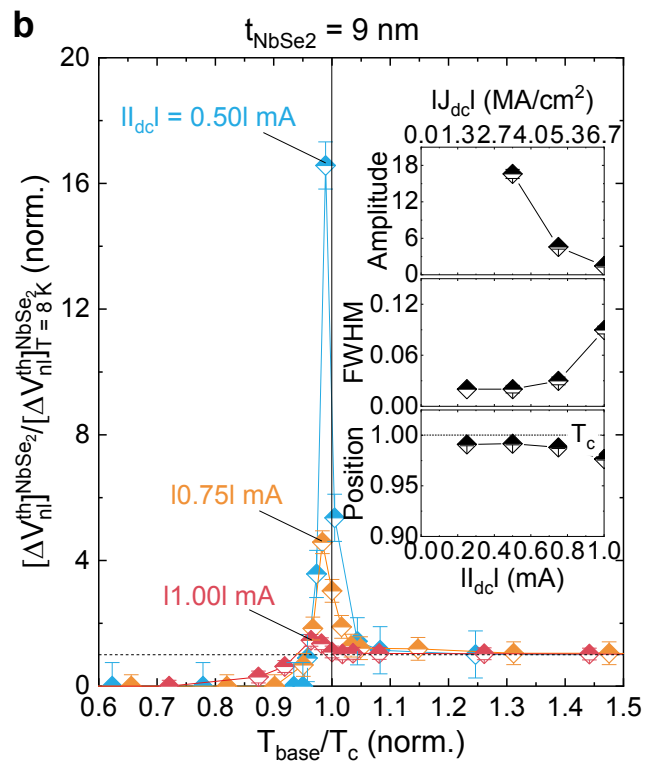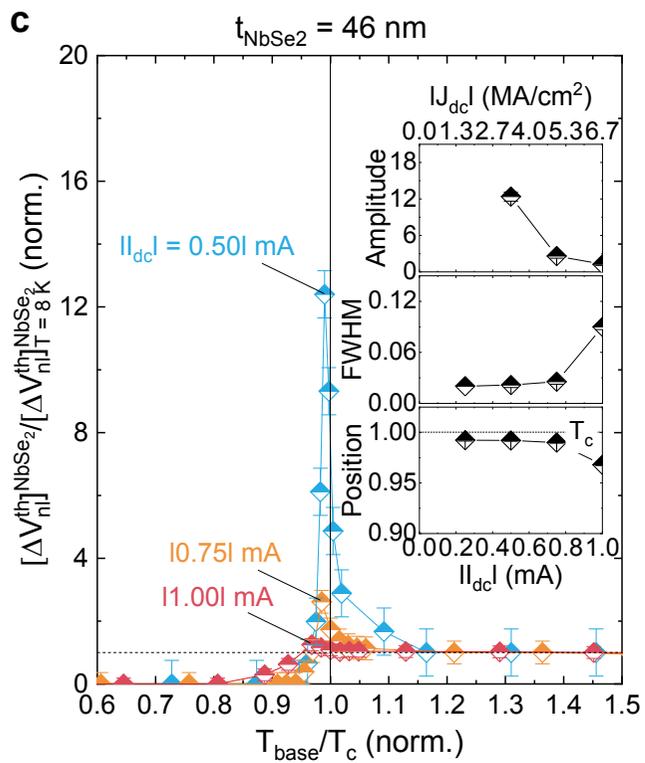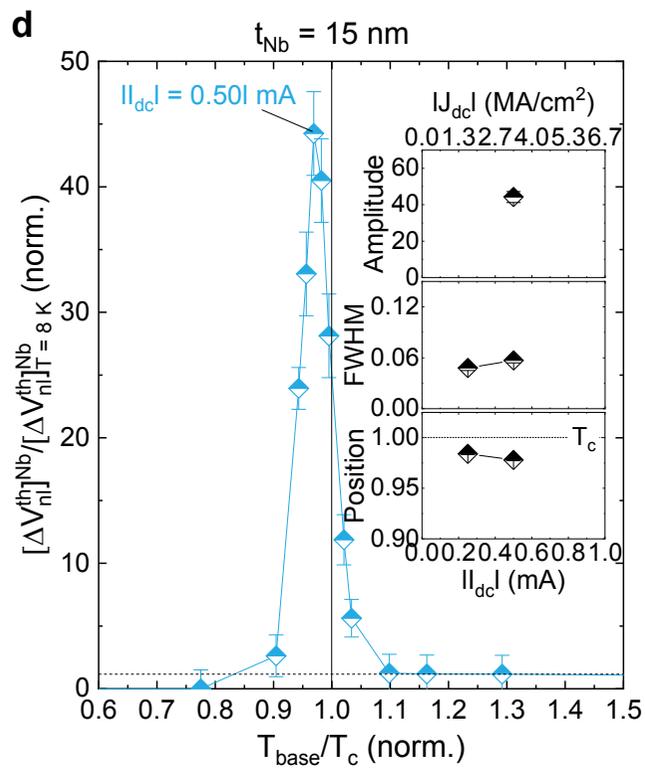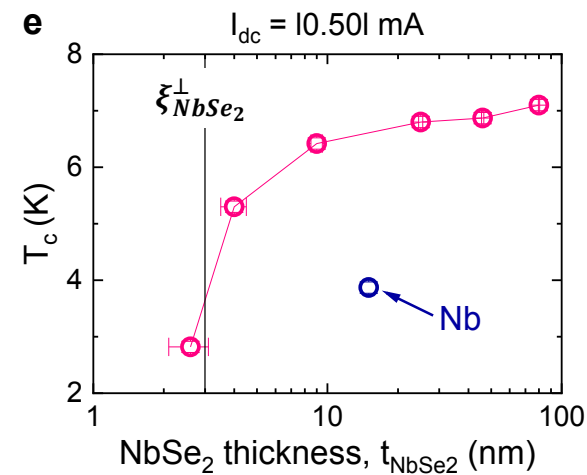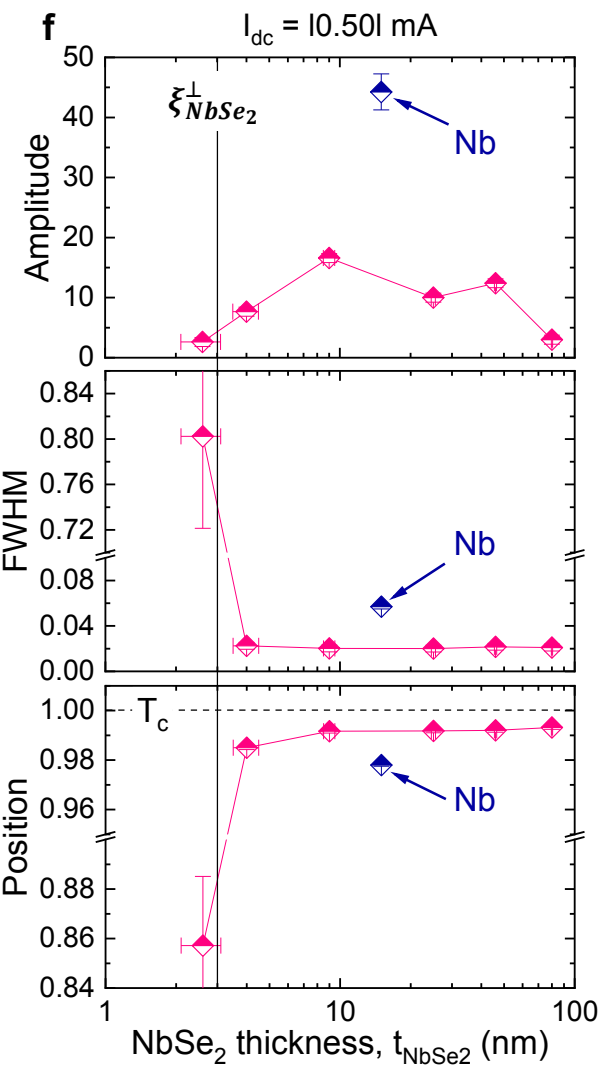

Supplementary Information

# Role of Two-Dimensional Ising Superconductivity in the Non-Equilibrium Quasiparticle Spin-to-Charge Conversion Efficiency

Kun-Rok Jeon,*† Kyungjune Cho,† Anirban Chakraborty, Jae-Chun Jeon, Jiho Yoon,

Hyeon Han, Jae-Keun Kim and Stuart S. P. Parkin*

*Max Planck Institute of Microstructure Physics, Weinberg 2, 06120 Halle (Saale), Germany*

†These authors contributed equally to this work.

†These authors contributed equally to this work.

*To whom correspondence should be addressed: jeonkunrok@gmail.com,

stuart.parkin@halle-mpi.mpg.de

**This PDF file includes:**

Supplementary Text

Figs. *S1* to *S3*

References (*S1-S9*)

**S1. Dry transfer of 2H-NbSe$_2$ flakes onto magnon spin-transport devices.**

In this section, we describe the device fabrication process (Figure S1) with a focus on dry transfer techniques[S1] used.

1) As outlined in Methods (main text), we first fabricated standard magnon device structures consisting of a YIG channel and two Pt electrodes, along with outer Au/Ru leads for electrical contacts to the central 2H-NbSe$_2$ flake.

2),3) To dry-transfer the 2H-NbSe$_2$ flake, we utilized a polypropylene carbonate (PPC) coated dome shaped polydimethylsiloxane (PDMS) stamp. To make the dome shaped PDMS stamp, the PDMS was mixed with curing agent in a ratio of 1:10 (curing agent: PDMS) and baked at 100 °C on SiO$_2$/Si wafer. The baked PDMS was treated with oxygen plasma for 5 min and a PPC layer was spin-coated on the PDMS. Subsequently, the PPC coated PDMS was transferred onto the glass slide. After we aligned the PDMS stamp to a suitable 2H-NbSe$_2$ which was pre-transferred onto a SiO$_2$(300 nm)/Si substrate using a micromanipulator, the PDMS stamp was slowly approached to the 2H-NbSe$_2$ flake and heated up to 100 °C. Next, we cooled the stamp down to 40 °C and picked the 2H-NbSe$_2$ flake up.

4),5) The picked-up 2H-NbSe$_2$ flake was aligned to the pre-patterned magnon device structure by the micromanipulator and released at 100 °C.

6) The remained PPC residue was removed by acetone.

7) To prevent the unintentional contribution of iSHE from inner Au/Ru leads themselves to total voltage signals, we electrically isolate them from the active regime of magnon spin-transport by depositing an Al$_2$O$_3$ oxide in-between apart from the electric contact parts on top of the central/transferred 2H-NbSe$_2$ flake.

8) Finally, we defined the inner Au/Ru leads.

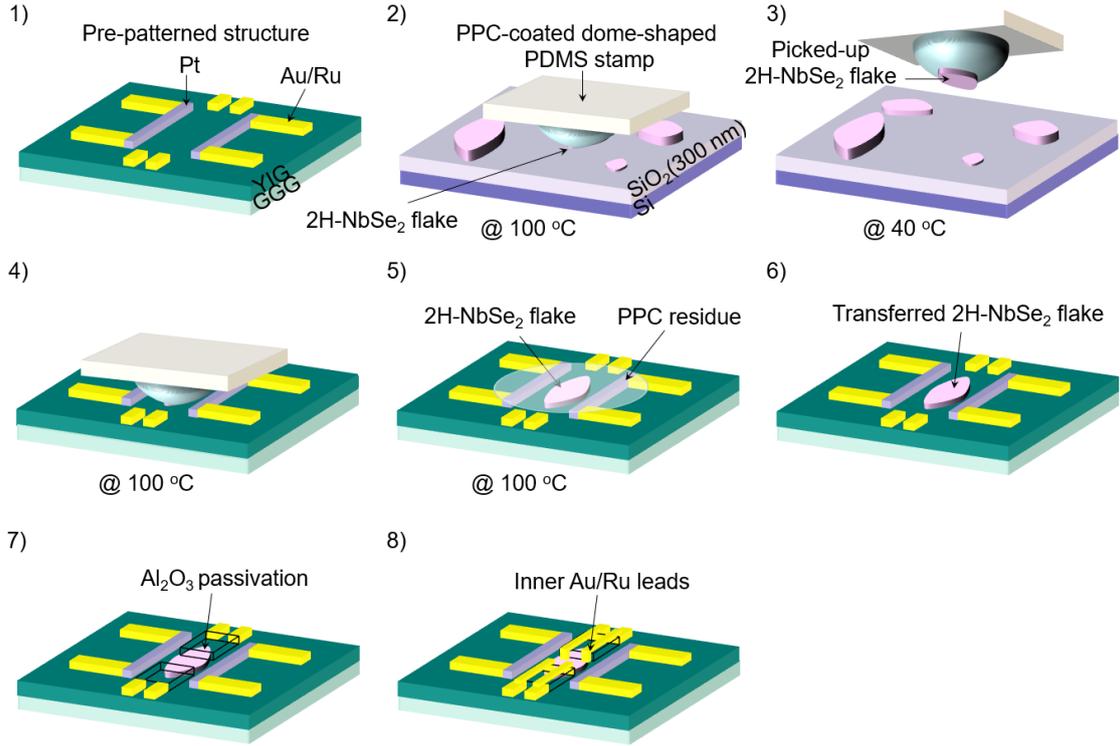

**Figure S1.** 1)-8) Schematic illustration of the device fabrication process flow.

**S2. Non-local spin signals detected by the Pt detector across $T_c$ of the 2H-NbSe$_2$ flake.**

For a given d.c. current $I_{dc}$ in the right Pt injector (Figure 2a,e), we simultaneously measure non-local voltages $[V_{nl}^{Pt}(\alpha), V_{nl}^{NbSe_2 \ (or \ Nb)}(\alpha)]$ across the left Pt detector and the central 2H-NbSe$_2$ (or Nb) detector as a function of IP magnetic-field-angle $\alpha$. From this, we can confirm that magnon-carried spin currents propagate through a YIG channel to the left Pt detector, located farther away than the central 2H-NbSe$_2$ (or Nb) detector from the Pt injector, and also identify a sign of the spin-Hall angle $\theta_{SH}$ for each detector. Given $\left[\Delta V_{nl}^{th}\right]^{Pt} > 0$ (Figure S2c,S2g), $\left[\Delta V_{nl}^{th}\right]^{NbSe_2} < 0$ (Figure S2d) and $\left[\Delta V_{nl}^{th}\right]^{Nb} < 0$ (Figure S2h), and the fact that Pt detector is well known to have a positive $\theta_{SH}$ (> 0),[S2] one can conclude that the $\theta_{SH}$ signs for 2H-NbSe$_2$ and Nb[S3-S5] detectors are both negative (< 0) and the 4d heavy element Nb of the 2H-NbSe$_2$ dominates spin-to-charge conversion phenomena.

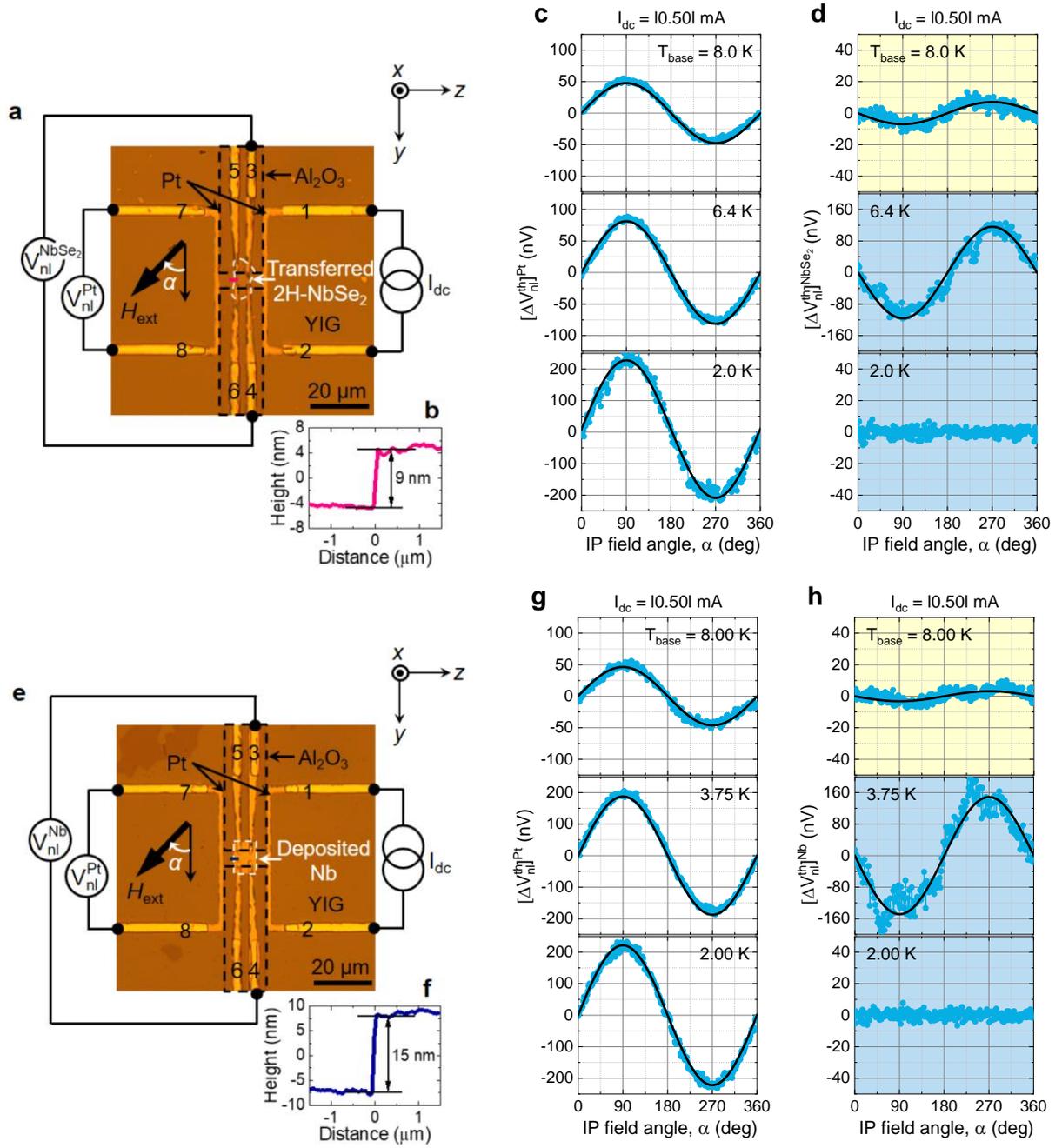

**Figure S2. a)** ptical micrograph of the $t_{NbSe_2} = 9$ nm device. **b)** Atomic force microscopy (AFM) scan of the transferred 2H-NbSe$_2$ ($t_{NbSe_2} = 9$ nm) flake. Thermally driven non-local voltages $\left[\Delta V_{nl}^{th}(\alpha)\right]^{Pt}$ (**c**) and $\left[\Delta V_{nl}^{th}(\alpha)\right]^{NbSe_2}$ (**d**) as a function of IP field angle $\alpha$ for the $t_{NbSe_2} = 9$ nm device, taken at $I_{dc} = |0.5|$ mA around $T_c$ of the 2H-NbSe$_2$. **e-h)** Data equivalent to **a-d** but for the $t_{Nb} = 15$ nm reference device. Note that except for **c** and **g**, the others are also presented in the main text (Figure 1,3).

## S3. Transition-state enhancement of QP iSHE for the $t_{NbSe_2}$ = 2.5 nm device.

As discussed in the main text (Figure 4e,f), for the $t_{NbSe_2}$ < 3 nm device, $T_c$ drops significantly and its transition width broadens anomalously (see Figure S3e). This causes sudden changes of the peak width and position of the transition-state QP iSHE (Figure 4f, main text) relative to the $t_{NbSe_2}$ > 3 nm devices. These results can be ascribed to depressed superconductivity and smearing-out effect of QP DOS around the gap edge[S5,S6] due to enhanced thermal fluctuations at the 2D limit[S7,S8] ($t_{NbSe_2} < \xi_{NbSe_2}^{\perp} \approx$ 3 nm), which inactivates the associated resonant absorption of magnon spin currents.[S4,S5,S9]

We also note that in the normal state ($T_{base} > T_c$), $\left[\Delta V_{nl}^{th}\right]^{NbSe_2}$ for $t_{NbSe_2}$ = 2.5 nm device (Figure S3d) is *at least 4 times* larger than $\left[\Delta V_{nl}^{th}\right]^{Nb}$ for $t_{Nb}$ = 15 nm device (Figure S2h), indicating high spin mixing conductance and spin transparency at the interface of our transferred 2H-NbSe$_2$ flake and YIG film.

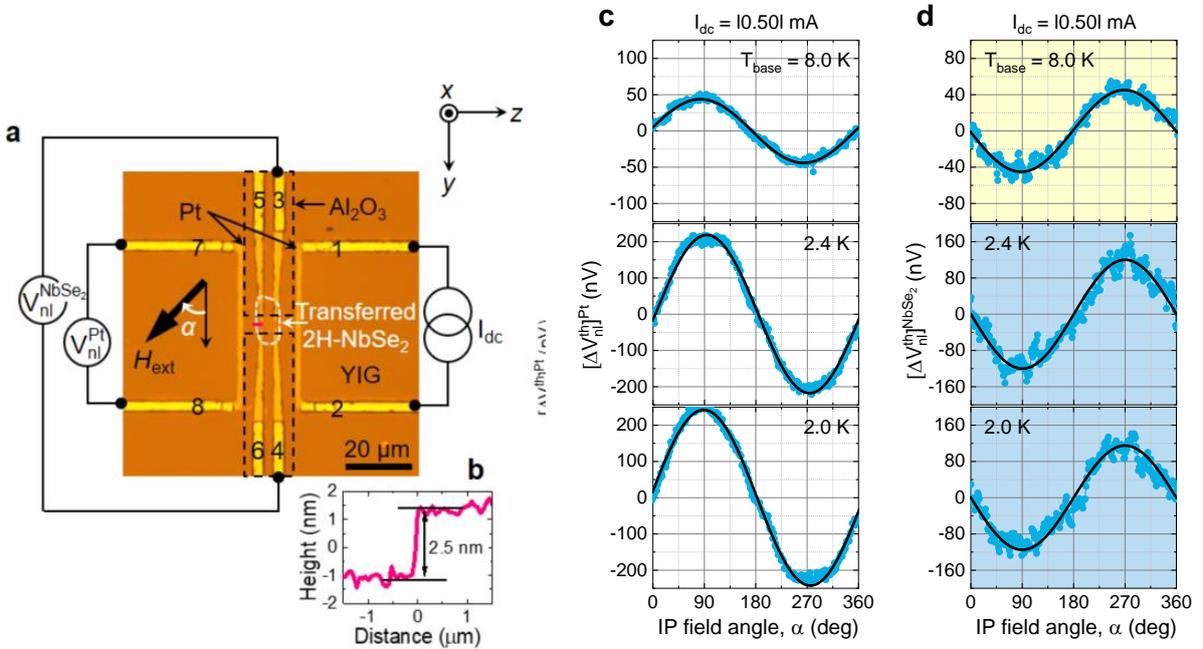

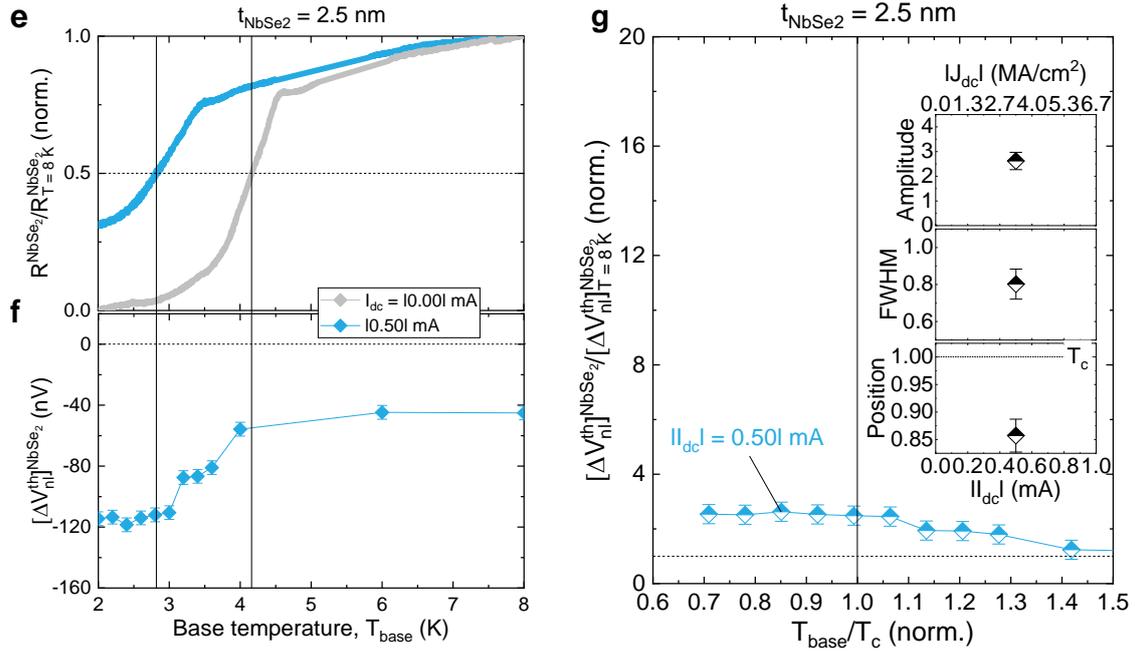

**Figure S3.** a) Optical micrograph of the $t_{NbSe_2}$ = 2.5 nm device. b) Atomic force microscopy (AFM) scan of the transferred 2H-NbSe$_2$ ($t_{NbSe_2}$ = 2.5 nm) flake. Thermally driven non-local voltages $[\Delta V_{nl}^{th}(\alpha)]^{Pt}$ (c) and $[\Delta V_{nl}^{th}(\alpha)]^{NbSe_2}$ (d) as a function of IP field angle $\alpha$ for the $t_{NbSe_2}$ = 2.5 nm device, taken at $I_{dc}$ = |0.5| mA around $T_c$ of the 2H-NbSe$_2$. e, Normalized 2H-NbSe$_2$ resistance $R^{NbSe_2}/R_{T=8K}^{NbSe_2}$ versus $T_{base}$ plots for the $t_{NbSe_2}$ = 2.5 nm device, measured using a four-terminal current-voltage method (using leads 3, 4, 5, 6 in **a**) for $I_{dc}$ = |0.0| and |0.5| mA in the Pt injector. The critical temperature $T_c$ is defined as the point where $R^{NbSe_2} = 0.5 R_{T=8K}^{NbSe_2}$. **f**) Estimated magnitude of $[\Delta V_{nl}^{th}]^{NbSe_2}$ as a function of $T_{base}$ for the $t_{NbSe_2}$ = 2.5 nm device. **g**) $[\Delta V_{nl}^{th}]^{NbSe_2}/[\Delta V_{nl}^{th}]_{T=8K}^{NbSe_2}$ versus $T_{base}/T_c$ plot for the $t_{NbSe_2}$ = 2.5 nm device. Each inset displays the $|I_{dc}|$ (or $|J_{dc}|$) dependence of the peak amplitude, width, and position.

## REFERENCES

[S1] R. Frisenda, E. Navarro-Moratalla, P. Gant, D. P. De Lara, P. Jarillo-Herrero, R. V. Gorbachev, A. Castellanos-Gomez, Recent Progress in the Assembly of Nanodevices and van der Waals Heterostructures by Deterministic Placement of 2D Materials. *Chem. Soc. Rev.* **2018**, *47*, 53.